------------------------------------------------------------------------
\documentstyle[preprint,aps]{revtex}
\begin{document}
\title{String Representation of Quantum Loops}
\author{Stefano Ansoldi \footnote{E-mail 
address: ansoldi@trieste.infn.it}}
\address{Dipartimento di Fisica Teorica\break
Universit\`a di Trieste,\\
INFN, Sezione di Trieste}
\author{Carlos Castro\footnote{E-mail 
address:castro@hubble.cau.edu}}
\address{
Center for Theoretical Studies of Physical Systems,\\
Clark Atlanta University,\\
Atlanta, Georgia 30314 USA}

\author{Euro Spallucci\footnote{E-mail 
address: spallucci@trieste.infn.it}}
\address{Dipartimento di Fisica Teorica\break
Universit\`a di Trieste,\\
INFN, Sezione di Trieste}
\maketitle
\begin{abstract}
We recover a general representation for the quantum state of a relativistic
closed line ({\it loop}) in terms of string degrees of freedom.
The general form of the loop functional splits into the product
of the {\it Eguchi functional,} encoding the {\it holographic quantum
dynamics}, times the Polyakov path integral, taking into account the 
full {\it Bulk dynamics,} times a {\it loop effective action,} which is 
needed to renormalize boundary ultraviolet divergences. 
The Polyakov string action is derived as an {\it effective action}
from a phase space, covariant, Schild action, by functionally integrating
out the world--sheet coordinates.  
The {\it area coordinates} description of the boundary quantum dynamics, 
is shown to be induced by the ``zero mode'' of the bulk quantum 
fluctuations. Finally, we briefly comment about a {\it ``unified,
fully covariant''} description of points, loops and strings in terms
of {\it Matrix Coordinates.}

\end{abstract}

\newpage

Non perturbative effects in quantum field theory are usually difficult
to study because of the missing of appropriate mathematical tools.
In the few cases where some special invariance, like electric/magnetic 
duality or super--symmetry, allows to open a window over the strong
coupling regime  of the theory, one  usually is faced with a new 
kind of solitonic excitations describing extended field configurations,
which are ``dual'' to the point--like states of the perturbative regime. 
Remarkable examples of extended structures of this sort ranges from the Dirac 
magnetic monopoles \cite{dirac} up to the loop states of quantum gravity
\cite{rovo}.\\
The dynamics of these extended objects is usually formulated in terms
of an ``effective theory'' of relativistic strings, i.e. one 
switches from a  quantum field theoretical framework to a 
different description by taking for granted that there is some
relation between the two. Such a kind of relation 
has  recently shown up in the context of super string theory as a web of 
dualities among different phases of different super string models. 
On the other hand,  there is still no explicit way to connect point like and 
stringy phases of non--supersymmetric field theories like QCD and
 quantum gravity.
\\
The main purpose of this letter is to recover  a general 
{\it String Representation}, of  a   loop wave functional in terms of 
the {\it Bulk} and {\it Boundary} wave functional of a quantum string.
The loop wave functional can describe the quantum state of
a one dimensional, closed, excitation of some quantum field theory, while
the corresponding string functional is induced by the quantum fluctuations
of an open world--sheet. We shall find a quite general relation linking
these two different objects.
The matching between a quantum loop and a quantum string being provided by 
a (functional) Fourier transform over an abelian vector field.\\ 
As a byproduct of this new {\it Loops/Strings connection}, we shall clarify
the interplay  between the bulk quantum dynamics, encoded into
 the Polyakov path integral, \cite{polya}, and the induced boundary dynamics.  
In this regard, Eguchi string quantum mechanics\cite{egu} will be shown
to provide a sort of {\it holographic } description of the boundary quantum
fluctuations.
\\
\\
The quantum state of a {\it closed, bosonic string}  is described by a
(complex) functional $\Psi[\,\overline y\,]$ 
 defined over the space of all possible configurations, i.e.
 {\it shape,} of the boundary $\overline y$ of an open 
 world--sheet $ Y$. Accordingly, $\Psi[\,\overline
 y\,]$ can be written as a phase space path integral of the form
\begin{equation}
\Psi[\,\overline y\,]= \int_{\partial Y= \overline y} 
[DY^\mu][DP_{m\,\mu}][Dg_{ab}]
\exp\left(i\int_\Sigma d^2\sigma\sqrt g \,L(Y\ , P\ , g_{ab}\ ;\sigma)\right)
\label{pathuno}
\end{equation}
where, $Y^\mu(\sigma)$ are the string coordinates mapping the string manifold
$\Sigma$ in a world surface in target space; 
$\sigma^m=\{\sigma^0\ , \sigma^1\}$ are internal coordinates,
and $g_{mn}(\sigma)$ is the metric  over the string manifold. We assume that
$\Sigma$ is simply connected and bounded by a curve $\partial\Sigma$
parametrically represented by $\sigma^m=\sigma^m(s)$. Then, $Y^\mu(\sigma)$
maps $\partial\Sigma$ in a contour $\partial Y\equiv \overline y$ in target 
space
\begin{equation}
Y^\mu\left(\sigma^0(s)\ ,  \sigma^1(s)\right)=\overline y^\mu(s)\ .
\end{equation}
$P_{m\,\mu}(\sigma)$ is the canonical momentum 
\begin{equation}
P^m{}_\mu(\sigma)\equiv {\partial L\over \partial \partial_m Y^\mu}\ .
\label{pcan}
\end{equation}
Finally, we sum in the path integral (\ref{pathuno}) over {\it all the
the string world sheets having the closed curve $x^\mu=\overline y^\mu(s)$ as
the only boundary.}\\
A closed line and an open two--surface can be given different
geometrical characterizations.
For our purposes, the best way to introduce them
is through the respective associated currents. Let  us define
 the  {\it Bulk Current} $J^{\mu\nu}(x\ ; Y)$, as a rank two
 antisymmetric tensor distribution having non vanishing
support over a two dimensional world surface, and the  {\it Loop Current}
$J^\mu(x\ ; C)$ as the vector distribution with support
over the closed line $x^\mu= x^\mu(s)$:
\begin{eqnarray}
&& J^{\mu\nu}(x\ ; Y)\equiv \int_Y dY^\mu\wedge dY^\nu\, \delta^D[x-Y]
\label{jbulk}\\
&& J^\mu(x\ ; C)\equiv \oint_C dx^\mu\, \delta^D[x-x(s)]\label{jloop}\ ,\\
&& J^\mu(x\ ; \overline y)\equiv \partial_\lambda J^{\lambda\mu}(x\ ; Y)
\label{jbordo}
\end{eqnarray}
The divergence of the bulk current define the {\it Boundary current}
(\ref{jbordo}).
A simple way to link a closed line to a surface is by ``appending'' 
the surface to the assigned loop \cite{axion}. 
This matching condition can be formally written by identifying the loop
current with the boundary current :  

\begin{equation}
J^\mu(x\ ;C)=J^\mu(x\ ; \overline y)=
\partial_\nu J^{\nu\mu}(x\ ; Y)\label{div}\ .
\end{equation}
It is worth to remark that the equation (\ref{div}) 
defines $J^\mu(x\ ; C)$ as the current associated to the
boundary of the  surface $Y$. In the absence of (\ref{div}) 
$J^\mu(x\ ; C)$ is a loop current with no reference to
any surface. In other words, equation (\ref{div}) is 
a formal description of the ``~{\it gluing operation }~'' between
the surface $Y$ and the closed line C.

Accordingly, one would relate loops states to string states through
a functional relation of the type
\begin{equation}
\Psi[\, C \,]=\int [\,d\,\overline y\,]\delta\left[\, C - \overline y\,\right]
\Psi[\,\overline y\,]
\label{lstate}
\end{equation}
where, we introduced a {\it Loop Dirac delta} function which picks up
the assigned loop configuration C among the (infinite) family of all
the allowed string boundary configurations $\{  \overline y^\mu(s) \}$.
Such a loop delta function requires a more appropriate definition. The
``current representation'' of extended objects provided by (\ref{jbulk}),
(\ref{jloop}), (\ref{jbordo}) offers a suitable ``Fourier'' form
of $\delta[x-\overline y]$ as a functional integral over a vector field
$A_\mu(x)$:

\begin{equation}
\delta\left[J^\mu(x\ ;C)- J^\mu(x\ ; \overline y)\right]
=\int [DA_\mu(x)]\exp\left\{-i\int d^D x A_\mu(x)\left[ 
J^\mu(x\ ; C)- J^\mu(x\ ;\overline y)\right]\right\}\ ,
\label{delta}
\end{equation}

and the loop functional (\ref{lstate})      can be written as
\begin{eqnarray}
\Psi[\, C \,]=&&\int [d\,\overline y^\nu] [DY^\mu][DP_{m\,\mu}][Dg_{ab}][DA_\mu]
\times\nonumber\\
&&\exp\left(-i\int d^Dx A_\mu(x)\left[\,
J^\mu(x\ ; C)- J^\mu(x\ ; \overline y)\,\right]
+i\int_\Sigma d^2\sigma\sqrt{g} \,L(\,Y^\mu\ , P^m{}_\mu\ , g_{ab}\ ;\sigma\,)
\right)
\label{pathdue}
\end{eqnarray}

 These seemingly harmless manipulations are definitely non trivial. 	
 A proper implementation of the boundary conditions 
 introduces an {\it abelian vector field } coupled both to the loop and 
 the boundary currents. The first integral in (\ref{delta}) is 
 the {\it circulation} of $A$ along the loop C while the second
 represents the circulation along $\overline y$ :
 \begin{eqnarray}
 &&  \int d^D x \,A_\mu(x)\,J^\mu(x\ ;C)=\oint_C dx ^\mu A_\mu(x)\ ,
 \label{circ1}\\
 &&  \int d^D x\, A_\mu(x)\,J^\mu(x\ ; \overline y)=
 \oint_\gamma d\,\overline y^\mu A_\mu(\overline y)
 \label{flux}
 \end{eqnarray}
 
 After introducing the Wilson factor as
 \begin{equation}
 W[\, A_\mu\ ,\gamma\,]\equiv \exp \left[
 (-i)\oint_{\gamma} dz^\mu A_\mu(z)\right]\label{wilson}
 \end{equation}
 the loop delta functional can be written as
 \begin{equation}
 \delta\left[\,C-\gamma\,\right]=\int D[\, A_\mu\,]W^{-1}[\, A_\mu\ , C\,] 
 W[\, A_\mu\ ,\gamma\,] 
 \end{equation}
 and we can define the {\it dual string functional} through 
 a {\it Loop Transform }\cite{book} :
 \begin{equation}
  \Phi[\, A_\mu(x)\,]\equiv \int [D\overline y^\nu]\, W[\, A_\mu\ ,\gamma\,]
  \,\Psi[\,\gamma\,]\ .
  \end{equation}
Hence, the vector field $A_\mu(x)$ is the Fourier conjugate variable to
the string boundary configuration. Finally, by projecting $\Phi[\, A_\mu(x)\,]$
along the loop C we obtain the wanted result
\begin{eqnarray}
 \Psi[\, C \,] =\int [DA_\mu] W^{-1}[\,A_\mu\ ,C\,]\Phi[\,A_\mu\,]
\label{looptransf}
\end{eqnarray}
The whole procedure can be summarized as follows:
\begin{eqnarray}
 &&{\cal L}_A\{\hbox{string functional}\}\longrightarrow  
 \hbox{dual string functional}\\
 &&{\cal L}_C^{-1} \{     \hbox{dual string functional} \}  
 \longrightarrow  \hbox{loop functional}
  \end{eqnarray}

Let us proceed by unraveling the information contained in the string
functional $\Psi[\,\gamma\,]$. First of all, we must
choose the classical string action
and pushing forward the functional integration. There are several 
action functionals providing equivalent classical descriptions of
string dynamics: the most appropriate for our purpose is the {\it
``covariant'' Schild action:}
\begin{eqnarray}
S[\, Y^\mu\ , P^m{}_\mu\ , g_{mn}\ ;\sigma\,]&=
&\int_\Sigma d^2\sigma\,\sqrt g \,
\partial_n Y^\nu\,P^n{}_\nu
-{1 \over 2\mu_0 }\int_\Sigma d^2\sigma\sqrt g\, g^{ab}\, P_{a\,\mu}P_b{}^\mu
\nonumber\\
&=& {1\over 2}\int_Y dY^\mu\wedge  dY^\nu P_{\mu\nu}
-{1\over 2\mu_0}\int_\Sigma d^2\sigma\sqrt g \,
g^{ab}\, P_{a\,\mu}P_b{}^\mu\nonumber\\
\label{covact}
\end{eqnarray}
where, $\mu_0=1/2\pi\alpha^\prime$ is the string tension, and $P_{\mu\nu}$
is the area momentum \cite{noi} conjugated to world--sheet tangent bi--vector
\footnote{We indicate with $\delta^{[\,m\,n\,]}$ the antisymmetric symbol
while $ \epsilon^ {m\,n} \equiv \delta^{[\,m\,n\,]}/\sqrt g $ is the
Levi--Civita tensor.}:
\begin{eqnarray}
P_{\mu\nu}\equiv {\partial L\over \partial Y^{\mu\nu}}\ ,\quad 
Y^{\mu\nu}\equiv \epsilon^{m\,n}\partial_{[\,m} Y^\mu\partial_{n\,]}
Y^\nu\ .\label{areamom}
\end{eqnarray}
 
 Variation of the action functional with respect the field variables
 provides the ``classical equation of motion''\footnote{Note that
 the first term in (\ref{covact}) is independent from the string
 metric $g_{mn}$. Only the second term contributes to the string
 energy momentum tensor.}
 \begin{eqnarray}
 &&\epsilon^{m\,n}\partial_{[\,m}\,P_{n\,]\mu}=0\ ,\label{uno}\\
 &&P_{m\,\mu}=P_{\mu\nu} \,\epsilon_m{}^n \partial_n Y^\nu\ ,\label{due}\\
 &&-P_m{}^\mu P_n{}_\mu + {1\over 2}g_{mn}\, g^{ab}P_a{}^\mu P_b{}_\mu=0
\ .\label{tre}
 \end{eqnarray}
 Equation (\ref{tre}) requires the vanishing of the string energy--momentum
 tensor $T_{mn}$. Eq.(\ref{due}) allows us to solve eq.(\ref{tre}) with
 respect to the string metric:
 \begin{equation}
 g_{mn}=\partial_m Y^\mu\, \partial_n Y_\mu \label{gonn}\equiv\gamma_{mn}(Y)
 \end {equation}
 Eq.(\ref{due}) and (\ref{gonn})
 show that the on--shell canonical momentum is proportional
 to the gradient of the string coordinate and the on--shell string metric
 matches the world--sheet induced metric. By inserting these classical
 solutions into (\ref{covact}) one recovers the Nambu--Goto action:
 \begin{equation}
 S[\,Y^\mu(\sigma)\ , P_{\mu\nu}\ ,\gamma_{mn}(Y)\ ;\sigma\,]=
 -\mu_0\int_\Sigma d^2\sigma\sqrt{\vert det\,[\gamma_{mn}(Y)]\vert}\ .
 \label{ng}
 \end {equation}
 With some hindsight, this result follows from having introduced a non--trivial
 metric $g_{mn}(\sigma)$ in the string manifold. Thus, the Schild action becomes
 diffeormophism invariant as the Nambu--Goto action. Accordingly, we recovered
 the classical equivalence showed in (\ref{ng}).
 To carry on the path integration it is instrumental to extract
  a pure boundary term from  the first integral in (\ref{covact}) 
 \begin{eqnarray}
 {1\over 2}\int_Y dY^\mu\wedge  dY^\nu P_{\mu\nu} &=& 
 {1\over 2}\int_Y d \left( Y^\mu  dY^\nu P_{\mu\nu}\right)-
 {1\over 2}\int_Y Y^\mu  dP_{\mu\nu}  \wedge  dY^\nu \nonumber\\
 &=& {1\over 2}\oint_\gamma
 \overline y^\mu  d  \overline y^\nu P_{\mu\nu}(\overline y)-
 {1\over 2}\int_\Sigma d^2\sigma  Y^\mu(\sigma)  
 \epsilon^{mn} \partial_{[\,m} P_{n\,]\mu}  \label{phase}
 \end{eqnarray}
 Then, we recognize that $Y^\mu(\sigma)$ appears in the path integral
 only in the last term of (\ref{phase}) through a linear coupling to the 
 left hand side of the classical equation of motion  (\ref{uno}). 
 Accordingly, to integrate over the string coordinate is tantamount to
 integrate over a {\it Lagrange multiplier} enforcing the canonical
 momentum to satisfy the classical equation of motion (\ref{uno}):
 \begin{equation}
 \int [DY^\mu(\sigma)]\exp\left[{1\over 2}\int_\Sigma d^2\sigma 
 Y^\mu(\sigma)\,\epsilon^{mn} \partial_{[\,m} P_{n\,]\mu}\right]=
 \delta\left[ \partial_{[\,m} P_{n\,]\mu}\right]
 \end{equation}
 Once the string coordinates have been integrated out, the resulting
 path integral reads
 
 \begin{equation}
 \Psi[\,\gamma\,]=
 \int [Dg_{mn}][DP_{m\mu}]\delta\left[ \partial_{[\,m}
 P_{n\,]\mu}\right]\exp\left({i\over 2}
 \oint_\gamma\overline y^\mu  d\,\overline y^\nu P_{\mu\nu}
 (\overline y)-{i\over 2\mu_0}
 \int_\Sigma d^2\sigma\sqrt g\, g^{mn} P_m{}_\mu P_n{}^\mu\right)
 \ .\label{pathcl}
 \end{equation}
 Eq.(\ref{pathcl}) shows that we have sum only over classical momentum
 trajectories. Such a restricted integration measure span the subset of 
 momentum  trajectories of the form
 \begin{eqnarray}
 &&P_{\mu\nu}=\overline  P_{\mu\nu}+\sqrt{\mu_0}\, Q_{\mu\nu}(\sigma)\\
 &&P_m{}_\mu(\sigma)=\overline P_{\mu\nu}\,\epsilon_m{}^n
 \partial_n Y^\nu (\sigma)
 +\sqrt{\mu_0}\,\partial_m\eta_\mu(\sigma)\ ,\quad  \epsilon_{mn} Q_{\mu\nu}
 \partial^n Y^\nu=
 \partial_m\eta_\mu
 \label{trial}
 \end{eqnarray}
 where, $\eta_\mu(\sigma)$ is a $D$--components multiplet of world--sheet 
 scalar fields, and $ \overline P_{\mu\nu} $ is a constant background over 
 the string manifold, i.e.$  P_{\mu\nu} $ is the {\it area momentum zero mode :}
 \begin{equation}   
 \partial_m \overline P_{\mu\nu}=0\ .
  \end{equation}
  By averaging the  $\eta$--field over the string world--sheet, one can extract
  its zero frequency component $ \overline\eta^\mu $   : 
  \begin{equation}
  \eta^\mu(\sigma)= \overline\eta^\mu +\widetilde \eta^\mu(\sigma)\ ,\quad
  \overline\eta^\mu\equiv {1\over \int_\Sigma d^2\sigma\sqrt g}\int_\Sigma
  d^2\sigma\sqrt g \,\eta^\mu(\sigma)
  \end{equation}
  $\widetilde \eta^\mu(\sigma)$ describes the {\it bulk quantum fluctuations,} 
  as measured with
  respect to the reference value $ \overline\eta^\mu $. Confining \footnote{
  By ``{\it confinement}'' we mean that there is no leakage of the field
  current $J_m\equiv \widetilde\eta^\mu\partial_m \widetilde\eta_\mu$ 
  off the world sheet boundary,i.e.
  $$
  \oint_\gamma dn^a J_a=0\ .
  $$}
  $\widetilde \eta^\mu(\sigma)$
  to the bulk of the string world--sheet requires appropriate boundary
  conditions. Accordingly,
  we assume  that both $ \widetilde\eta^\mu(\sigma)$ and its 
  (normal and tangential) 
  derivatives vanish when restricted on the boundary $\gamma$:
 
 \begin{eqnarray}
 &&\widetilde\eta^\mu\vert_\gamma=0\ ,\label{b1}\\
 &&t^m\partial_m   \widetilde\eta^\mu\vert_\gamma =0\ ,\label{b2}\\
 &&n^m\partial_m  \widetilde\eta^\mu\vert_\gamma=0\ .\label{b3}
 \end{eqnarray}

  Now, we can give a definite meaning to the integration measure over 
  the classical solutions:
  \begin{equation}
 \int [DP_{m\mu}]\delta\left[ \partial_{[\,m}
 P_{n\,]\mu}\right]= \int d^D\overline\eta\int [d\,\overline P_{\mu\nu}]
 \int [{\cal D}\widetilde\eta_\mu(\sigma)]\ ,\\
 \end{equation}
 
 We remark that the first two integrations are ``over numbers'' and not over 
 functions.
 We have to sum over all possible constant values of $\overline P_{\mu\nu}$ and
 and $\overline\eta^\mu$. 
 The constant mode of the bulk momentum does not mix with the other modes
 in the on--shell Hamiltonian because the cross term vanish identically
 \begin{equation}
 \overline P_{\,[\mu\nu\,]}g^{(m\,n)}\,\partial_{\,(m }Y^{\,[\nu}
  \,\partial_{n)}\eta^{\mu\,]}\equiv 0\ ,\qquad \delta^{[mn]}\partial_{[m}
  Y^\mu\partial_{n]}\eta_\mu\equiv 0\ .
 \end{equation}
 
 Accordingly, boundary dynamics decouples from the bulk dynamics\footnote{
 A boundary term
 $$
 {1\over 2\sqrt{\mu_0}}\overline P_{\mu\nu}\oint_\gamma
 dt^m\,\eta^{[\,\mu}
 \,\partial_m Y^{\nu\,]} + {1\over 2}
 \oint_\gamma dn^m \, \eta^\mu\,\partial_m\,\eta_\mu
 $$
 vanishes because of the boundary conditions (\ref{b1}),(\ref{b2}), (\ref{b3}).}
 :
 
 \begin{eqnarray}
 {1\over 2}
 \oint_\gamma\overline y^\mu  d  \overline y^\nu P_{\mu\nu}
 (\overline y)&=&{1\over 4}\overline P_{\mu\nu}
 \oint_\gamma d\sigma^m\,\overline y^{[\,\mu}\, \partial_m  \overline y^{\nu\,]}
 \equiv {1\over 2}\overline P_{\mu\nu}\,\sigma^{\mu\nu}(\gamma)
 \label{bt}\\
 -{1\over 2\mu_0}\int_\Sigma d^2\sigma\sqrt g g^{mn} P_m{}_\mu P_n{}^\mu &=&
 -{1\over 4\mu_0}\overline P_{\mu\nu} \,  \overline P^{\mu\nu}\int_\Sigma 
d^2\sigma
 \sqrt g  - {1\over 2}\int_\Sigma d^2\sigma\sqrt g\,
 \eta_\mu\,\Delta_g \eta^\mu 
 \label{honn}\ ,
 \end{eqnarray}
 where, $\Delta_g$ is the covariant, world sheet, D'Alembertian.
  We introduced in eq.(\ref{bt}) and  (\ref{honn})  two important ``areas''. 
 First, the loop  area tensor, or Pl\"ucker coordinate, 
 \begin{equation}
 \sigma^{\mu\nu}(\gamma)={1\over 2}\oint_{\gamma}
 \left( \overline y^\mu d\,\overline 
 y^\nu -    \overline y^\nu d\,\overline y^\mu\right)
 \ , \label{plucker}
 \end{equation}
 appears as the canonical partner of the zero mode bulk area momentum
 $\overline P_{\mu\nu}$. Each component of the $\sigma^{\mu\nu}(\overline y)$
 represents the area of the ``loop shadow'' over the $(\mu-\nu)$
 coordinate plane. These shadows are the two dimensional pictures of the
 string boundary, and  provide an ``{\it holographic
 coordinate system}''. The images allows to reconstruct the shape of the
 loop in a similar way an hologram encodes on a plate
 the whole information about a three dimensional structure.\\
 Second, the {\it proper area} of the string world--sheet
 \begin{equation}
 \int_\Sigma d^2\sigma\sqrt g\equiv A\label{area}
  \end{equation}
  provides an intrinsic evolution parameter for the system. However, 
  a quantum world sheet has not a definite proper area, the metric 
  in (\ref{area}) being a quantum operator itself. Thus, the l.h.s.
  of the definition (\ref{area}) has to be replaced by the
  corresponding quantum expectation value. 
  Then, we can split the sum over the string metrics into
   a sum over metrics $h_{mn}$, with fixed quantum expectation value of the 
   proper area, times an ordinary integral over all the values of the area
   quantum average:
  \begin{equation}
  \int [Dg_{mn}(\sigma)]\left(\dots\right)=
  \int_0^\infty dA \exp\left(i\lambda A\right) \int [Dh_{mn}(\sigma)]
  \exp\left(-i\lambda\int_\Sigma d^2\sigma \sqrt{h}\right)\left(\dots\right)\ .
  \end{equation}
  The $\lambda$ parameter enters the path integral as a constant external
  source enforcing the condition that the quantum average
  of the proper area operator is $A$. From a physical point of view it
  represents the world sheet cosmological constant, or vacuum energy
  density.\\
  Hence, we can write the string functional as
  
 \begin{eqnarray}
 &&\Psi[\,\gamma\, ]=\int d^D\overline\eta
 \int_0^\infty dA\,\exp\left(i\lambda A\right)
 \int [d\,\overline P_{m\mu}]\int [Dh_{mn}(\sigma)]\,
 [{\cal D}\widetilde\eta^\mu(\sigma)]\times\nonumber\\
 &&\exp\left({i\over 2}\overline P_{\mu\nu}\, \sigma^{\mu\nu}(\gamma)
 -{i\over 4\mu_0}\overline P_{\mu\nu} \overline P^{\mu\nu}\, A\right)
 \times\nonumber\\
 &&\exp\left( -{i\over 2}\int_\Sigma d^2\sigma\sqrt h\, 
 \widetilde\eta_\mu\,\Delta_g \widetilde\eta^\mu -i\lambda \int_\Sigma 
d^2\sigma\sqrt h \right)
 \ .\label{pathonn}
 \end{eqnarray}
  The world sheet scalar field theory contains geometry
  dependent,  ultraviolet divergent quantities. This two dimensional
  quantum field theory on a Riemannian manifold can be
  renormalized by introducing suitable bulk and boundary ``counter terms''.
  These new terms absorb the ultraviolet divergencies, and
  represent induced weight factors in the functional integration over 
  the metric, $Dh_{mn}$\footnote{This functional integration measure has to be
   factored out by the orbit of the diffeomorphism group, and in the
   critical dimension by the Weyl group, as well.}, 
   and  the boundary shape $D\,\overline y$.\\
 The basic result we get from
  of the above calculation is to ``factorize''  boundary and bulk
 quantum dynamics as follows:
 \begin{equation}
 \Psi[\,\gamma\,]\equiv \Psi[\,\overline y^\mu(s)\ ,
 \sigma^{\mu\nu}(\gamma)\,]=\int d^D\overline\eta
 \left[\exp\left(iS^{eff}[\,\gamma\,]\right)
 \right]\int_0^\infty dA \,\exp\left(i\lambda A\right)
 \Psi[\,\sigma(\gamma)\ ; A]\, Z_{BULK}^A\ .\label{main}
 \end{equation}
 The bulk quantum physics is encoded into the Polyakov 
 partition function\cite{polya}
 \begin{equation}
  Z_{BULK}^A= \int[Dh_{mn}(\sigma)][{\cal D}\widetilde\eta_\mu(\sigma)]
 \exp\left[-{i\over 2}\int_\Sigma d^2\sigma\sqrt h\, 
	\widetilde\eta_\mu\Delta_h \,\widetilde\eta^\mu 
	- i\int_\Sigma d^2\sigma\sqrt h\,
 \left(\kappa R +\lambda\right)\right]\label{polya}
 \end{equation}
 for a scalar field theory covariantly coupled to $2D$ gravity 
 on a disk\footnote{
 Our result refers to the disk topology. The extension to more
 complex world sheet topology is straightforward: the single, bulk,
  partition functional has to be replaced by a sum over definite genus 
  path integrals, 
  i.e. $  Z_{BULK}^A\longrightarrow \sum_g  Z_{BULK}^{(g),A}$. 
  This is the starting point for introducing topology changing
  quantum processes in the framework of string theory.}.\\ 
 \begin{equation}
 \Psi [\sigma^{\mu\nu}(\gamma)\ ; A]\equiv 
 \int[d\,\overline P_{\mu\nu}]\exp\left[
 {i\over 2}\overline P_{\mu\nu}\,\sigma^{\mu\nu}(\gamma)-i
 \left({\overline P_{\mu\nu} \,\overline P^{\mu\nu}\over 4\mu_0  }
 \right)\, A\right]
 \ ,\label{eguchi}\\
 \end{equation}
 is the Eguchi wave functional encoding the holographic quantum mechanics
 of the string boundary. Finally, $S^{eff}[\,\gamma\,]$ is the effective
 action induced by the quantum fluctuation of the string world sheet. It is
 a local quantity written in terms of the ``counterterms'' needed to cancel
 the boundary ultraviolet divergent terms. The required counterterms are
 proportional to the loop proper length and extrinsic curvature.\\ 
 The most part of current investigations in quantum string theory 
 start from the Polyakov path integral and elaborate string theory as
 scalar field theory defined over a Riemann surface. String perturbation
 theory come from this term as an {\it expansion in the genus} of the
 Riemann surface. Against
 this background,  we assumed {\it the phase space, covariant, path integral
 for the Schild string as the basic quantity} encoding the whole
 information about string quantum behavior, and we recovered the Polyakov
 ``partition functional''  as an {\it effective path integral,} after
 integrating out the string coordinates and factorizing out the boundary 
 dynamics. It is worth to recall that the Conformal Anomaly and the 
 {\it critical dimension} are encoded into the Polyakov path integral. 
 Accordingly, they are bulk effects. But, this is not the end of the story. 
 Our approach provides the boundary quantum dynamics as well.  
 The fluctuations of the  $\overline\eta^\mu$  field {\it induces}
 the non--local part of the effective action for $\overline y(s)$ , while
 the world sheet vibrations induce the local, geometry dependent, terms.
  Furthermore, the Eguchi string wave 
 functional $\Psi[\,\sigma; A]$  encodes the quantum holograpic dynamics of 
 the boundary in terms of area coordinates. $\Psi$ gives the  amplitude to 
 find a closed string, with area tensor $\sigma^{\mu\nu}$, as the only 
 boundary of a world--sheet of proper area $A$, as it was  originally 
 introduced  in the areal formulation of ``string  quantum mechanics''. 
 This overlooked approach implicitly broke the accepted ``dogma'' that string 
 theory is intrinsically a ``second quantized field theory'', which cannot be
 given  a first quantized, or quantum mechanical, formulation. 
 This claim is correct as far as it
 is referred to the infinite vibration modes of the world sheet bulk. On the
 other hand,  boundary vibrations are induced by the one dimensional field
 living on it and by the constant zero mode of the area momentum  
 $\overline P_{\mu\nu}$. From this viewpoint, the Eguchi approach appears as a
 sort of Mini Superspace approximation of the full string dynamics in
 momentum space: all the infinite bulk modes, except the constant one, 
 has been frozen out. 
 The ``field theory'' of this single mode collapses into a generalized
 ``quantum mechanics'' where both spatial and timelike coordinates are
 replaced by area tensor and scalar respectively:  
  \begin{equation}
  \hbox{second quantized bulk dynamics}\longrightarrow \hbox{first quantized
  boundary dynamics}\ .
  \end{equation}
   In the absence of external interactions, the 
   quantum state of the  ``free world sheet boundary''  is represented
   by a  Gaussian wave functional
  \begin{equation}
  \Psi [\sigma\ ; A]\propto \left({\mu_0\over A}\right)^{D(D-1)/4}
  \exp\left(-i\mu_0 {  \sigma^{\mu\nu}(\overline y)\sigma_{\mu\nu}(\overline y)
  \over 4A  }\right)
  \end{equation}
	solving the corresponding ``Schr\"odinger equation''\cite{noi2}:
	\begin{equation}
	-{1\over 4\mu_0 l_\gamma}\int  ds
	\sqrt{\overline y^{\prime\,2}(s)}
	{\delta^2  \Psi[\,\sigma^{\mu\nu}\ ; A]\over \delta\sigma^{\mu\nu}(s)
	\sigma_{\mu\nu}(s)}=
	i{\partial \over\partial A }\Psi[\,\sigma^{\mu\nu}\ ; A]
	\label{schrod}
	\end{equation}	
	where, $l_\gamma$ is the proper length 
	\begin{equation}
	l_\gamma\equiv \int ds\sqrt{\overline y^{\prime\,2}(s)}\ ,
	\end{equation}
	
	Equation (\ref{schrod}) encodes the boundary quantum behavior 
	irrespectively of the bulk dynamics: the only memory of the 
	world--sheet is only through its proper area A, no other information, 
	e.g. about topology of the string world sheet, is transferred to the 
	boundary. Our formulation enlights the 
	complementary role played by the bulk and boundary formulation of 
	strings quantum dynamics and the subtle interplay between the two. 
	The wave equation (\ref{schrod}) displays one of the most intriguing 
	aspects of Eguchi formulation: the role of spatial coordinates is 
	played by the area tensor while the area $A$ is the evolution parameter. 
	Such an unusual dynamics is now explained as an induced effect due
	to the bulk zero mode $\overline P^{\mu\nu}$. \\
	Finally, the loop transform connects the string boundary 
	$\overline y$ with the given loop C. The functional $\Psi[\, C \,]$ is
	obtained by  replacing everywhere $\overline y$ with C in 
	Eq.(\ref{main}).\\
	 The main result of this matching is:\\
	 {\it to ``dress''
	a bare quantum loop with the all degrees of freedom carried by a 
	quantum string.}\\
	\\
	We conclude this paper by speculating about a ``fully covariant''
	formulation of the boundary wave wave equation (\ref{schrod}) and its
	physical consequences.\\
	The spacelike and timelike 
	character of the two area  coordinates $\sigma^{\mu\nu}$ and $A$ shows
	up in the Schr\"odinger, ``non--relativistic'' form, of
	the wave equation (\ref{schrod}) which is second order in $\delta/
	\delta\sigma^{\mu\nu}(s)$, and first order in $\partial/\partial A$.
	Therefore, it is intriguing to
	ponder about the form and the meaning of the corresponding {\it
	Klein--Gordon} equation. The first step towards the ``relativistic''
	form of (\ref{schrod}) is to introduce an appropriate coordinate system
	where $\sigma^{\mu\nu}$ and $A$ can play a physically equivalent
	role. Let us introduce a {\it matrix coordinate} ${\bf X}^{MN}$  where
	certain components are $\sigma^{\mu\nu}$ and $A$. There is a 
	great freedom in the choice of ${\bf X}^{MN}$. However, an interesting
	possibility would be
	\begin{equation}
	{\bf X}^{MN}=\left(\begin{array}{cc}
	              \sqrt{\mu_0}\,\sigma^{\mu\nu} & 
	              \delta^{\mu\nu}x_\nu^{C.M.}\\
	              \delta^{\mu\nu}x_\nu^{C.M.} & \sqrt{\mu_0}\,A_{\mu\nu}
	              \end{array}\right)\ , \quad       
	A_{\mu\nu}=\left(\begin{array}{cc}
	                 \vec 0 & A\\
	                 -A 	&\vec 0
	                 \end{array}
	            \right)\label{matrix}
	\end{equation}
	where, we arranged the string center of mass coordinate $ x_\nu^{C.M.}$
	inside off--diagonal sub--matrices and build--up an antisymmetric
	proper area tensor in oder to endow $A$ with the same tensorial
	character as $\sigma^{\mu\nu}$. The {\it  string length
	scale,}$1/\sqrt{\mu_0}$, has been introduced to provide the block
	diagonal area sub--matrices  the coordinate canonical dimension
	of a {\it length,} in natural units.
	The most remarkable feature for
	this choice of    ${\bf X}^{MN}$ is that if we let      
	$\mu\ ,\nu$ to range
	over four values, then ${\bf X}^{MN}$ is an $8\times 8$ anti symmetric
	matrix with {\it eleven} independent entries. Inspired by the
	recent progress in non--commutative geometry \cite{landi}, where
	point coordinates are described by non commuting matrices,  we 
	associate  to each matrix  (\ref{matrix}) a representative point 
	in an eleven dimensional space which is the product of the 
	$4$--dimensional spacetime, times, the $6$--dimensional holographic 
	loop space, times, the  $1$--dimensional areal time axis. Eleven 
	dimensional spacetime is the proper arena of $M$--Theory \cite{mth}, 
	and we do not believe this is a mere coincidence.\\
	According with the assignment of the
	eleven entries in ${\bf X}^{MN}$ the corresponding
	``point'' can represent different physical objects:\\
	i) a point--like particle, $\{ \sigma^{\mu\nu}=0\ , A=0\ ,
	x_\nu^{C.M.}\}$ ;\\
	ii) a loop with center of mass in $x_\nu^{C.M.}$ and holographic
	coordinates $\sigma^{\mu\nu}$,  $\{ \sigma^{\mu\nu}(\gamma)\ , A=0\ ,
	x_\nu^{C.M.}\}$ ;\\
	iii) an open surface of proper area $A$, 
	boundary holographic coordinates $\sigma^{\mu\nu}$ and
	center of mass in $x_\nu^{C.M.}$,i.e. a {\it real string},
	 $\{ \sigma^{\mu\nu}(\gamma)\ , A\ ,x_\nu^{C.M.}\}$;\\
	iv) a closed surface of proper area $A$, i.e. a virtual string,
	$\{ \sigma^{\mu\nu}=0\ , A\ , x_\nu^{C.M.}\}$.\\

	It is appealing to conjecture that ``Special Relativity'' in this
	enlarged space will transform one of the above objects into another
	by a reference frame transformation! From this vantage viewpoint
	particles, loops, real and virtual strings would appear as the same
	basic object as viewed from different reference frames. Accordingly,
	a quantum field $\Phi({\bf X})$ would create and destroy the
	objects listed above, or, a more basic object encompassing all of them.
	A {\it unified quantum field theory of points loops and 
	strings,} and its relation, if any,  with $M$--Theory or
	non--commutative geometry, is an issue which
	deserves a throughly, future, investigation \cite{cc}.  
	
	\section{Aknowledgements}
	One of us, C.C. would like to thank the italian M.U.R.S.T
	for financial support, and the Department of Theoretical Physics 
	of the University of Trieste for its hospitality during the completion
	of this work.

	\end{document}